\documentclass[pra,aps,10pt,amsfonts,floatfix,superscriptaddress,longbibliography,twocolumn]{revtex4-1}
\usepackage{mathbbol}
\usepackage{graphicx,color}
\usepackage{amsmath,amssymb,bm,amsthm}
\usepackage{epstopdf}
\usepackage{mciteplus}
\usepackage{bbding}
\usepackage{float}

\def\be{\begin{equation}}
\def\ee{\end{equation}}
\def\ba{\begin{eqnarray}}
\def\ea{\end{eqnarray}}

\begin{document}
\title{Self-averaging in many-body quantum systems out of equilibrium:\\ Time dependence of distributions}
\author{E. Jonathan Torres-Herrera}
\author{Isa{\'i}as Vallejo-Fabila}
\affiliation{Instituto de F\'isica, Benem\'erita Universidad Aut\'onoma de Puebla,
Apt. Postal J-48, Puebla, 72570, Mexico}
\author{Andrei J. Mart{\'i}nez-Mendoza}
\affiliation{Divisi\'on de Estudios de Posgrado e Investigaci\'on, Tecnol\'ogico Nacional de M\'exico/ Instituto Tecnol\'ogico de Oaxaca, C. P. 68030, Oaxaca de Ju{\'a}rez, Mexico}
\author{Lea F. Santos}
\affiliation{Department of Physics, Yeshiva University, New York City, New York, 10016, USA}

\date{\today}

\begin{abstract}
In a disordered system, a quantity is self-averaging when the ratio between its variance for disorder realizations and the square of its mean decreases as the system size increases. Here, we consider a chaotic disordered many-body quantum system and search for a relationship between self-averaging behavior and the properties of the distributions over disorder realizations of various quantities and at different timescales. An exponential distribution, as found for the survival probability at long times,  explains its lack of self-averaging, since the mean and the dispersion are equal. Gaussian distributions, however, are obtained for both self-averaging and non-self-averaging quantities. Our studies show also that one can make conclusions about the self-averaging behavior of one quantity based on the distribution of another related quantity. This strategy allows for semianalytical results, and thus circumvents the limitations of numerical scaling analysis, which are restricted to few system sizes.
\end{abstract}

\maketitle
\section{Introduction}
\label{intro}
Experimental advances with cold atoms~\cite{Bernien2017}, ion traps~\cite{TanARXIV}, superconducting devices~\cite{Martinis2020}, and nuclear magnetic resonance platforms~\cite{Niknam2020,Sanchez2020} allow for the high level of control and long coherence times of many-body quantum systems. This has invigorated experimental and theoretical studies of the long-time evolution of these systems. Common questions include the viability of thermalization~\cite{Rigol2008,Torres2013,Borgonovi2016,Dalessio2016}, the description of the dynamics~\cite{Bloch2012,Gobert2005}, and the time to reach equilibrium~\cite{Schiulaz2019,Dymarsky2019}. Much less explored is the question of self-averaging~\cite{Schiulaz2020,Torres2020,Richter2020}. 

A quantity of a disordered system is self-averaging when its relative variance --- the ratio between its variance for disorder realizations and the square of its mean --- decreases as the system size increases. If self-averaging holds, as the system size increases, then one can decrease the number of samples used in theoretical and experimental analyses. In this case, the properties of the system do not depend on the specific realization selected. Lack of self-averaging, however, makes the study of disordered systems more challenging. Take as an example the scaling analysis of many-body quantum systems. The problem is already hard, because the many-body Hilbert space grows exponentially with system size. If in addition to this, one cannot decrease the number of disorder realizations as the system size grows, the problem becomes intractable. 

Non-self-averaging behavior is often associated with disordered many-body quantum systems at the  transition between the delocalized and the localized phase~\cite{Serbyn2017} and systems at a critical point in general~\cite{Binder1986,Wiseman1995,Aharony1996,Wiseman1998,Orlandini2002,Castellani2005,Malakis2006,Roy2006,Monthus2006,Efrat2014}. This sort of studies have mostly been done at equilibrium~\cite{Aharony1996}. Recently, however, the analysis has been extended to systems out of equilibrium close to the localization transition point~\cite{Torres2020,Richter2020} and also in the chaotic regime~\cite{Schiulaz2020}. It has been shown that self-averaging is not directly related with quantum chaos~\cite{Schiulaz2020,Prange1997,Kunz1999,Kunz2002}, as one might naively expect.

Quantum chaos refers to specific properties of the eigenvalues and eigenstates of systems that are chaotic in the classical limit. The eigenvalues are correlated~\cite{MehtaBook,HaakeBook,Guhr1998} and the eigenstates are close to the random vectors~\cite{Zyczkowski1990,Zelevinsky1996,Borgonovi2016} of full random matrices. If the system shows these properties, then it is usual to refer to it as chaotic even if its classical limit is not well defined.

In Ref.~\cite{Schiulaz2020}, the analysis of self-averaging was done for both a disordered spin model in the chaotic regime and a model consisting of full random matrices of a Gaussian orthogonal ensemble (GOE). It was shown numerically and analytically that the survival probability (the probability for finding the system in its initial state at a later time) is non-self-averaging at any timescale. Other quantities considered include the inverse participation ratio, which measures the spread of the initial state in the many-body Hilbert space, and observables measured in experiments with cold atoms and ion traps, namely the spin autocorrelation function and the connected spin-spin correlation function. The self-averaging behavior of the inverse participation ratio and spin autocorrelation function varies in time, while the connected spin-spin correlation function is self-averaging at all times.

Motivated by the results in Ref.~\cite{Schiulaz2020}, we now study numerically and analytically the distributions over disorder realizations of those same quantities throughout their evolution to equilibrium using again both the GOE and the disordered spin model. In addition, to avoid the negative values that can be reached with the  spin autocorrelation function, we consider also the absolute value and the square of the spin autocorrelation function.  Our goal is to understand how the shape and overall properties of the distributions depend on time, observables, and models, and whether they can help us determine when self-averaging holds. 

We find that at short times, the distributions are model dependent. Due to the locality of the spin model Hamiltonian, the distributions of the quantities considered here exhibit a fragmented structure with peaks at different energy windows, while the distributions are Gaussian for the GOE model. 

At long times, the distributions become similar for both models, but they differ depending on the quantity. The survival probability, for example, shows an exponential distribution~\cite{Argaman1993,Prange1997,Kunz1999,Kunz2002} as soon as the correlations between the eigenvalues get manifested in the dynamics. This distribution, where mean and standard deviation coincide, explains the lack of self-averaging of this quantity at long times. For the other quantities, the distribution is either Gaussian or related to a normal distribution. Gaussian distributions are found for both self-averaging and non-self-averaging quantities. 

A useful outcome of these studies is the realization that the shape of the distribution for one quantity can assist with the analysis of the self-averaging behavior of another related quantity. As an example, we discuss the case of the spin autocorrelation function, $I(t)$, and its absolute value, $|I(t)|$. The numerical analysis of the self-averaging behavior of $|I(t)|$ at long times are inconclusive, due to the limited system sizes available. However, in hands of the Gaussian distribution for $I(t)$, we find analytically the dependence on system size of the relative variance of $|I(t)|$. With this strategy, we are able to deduce that $|I(t)|$ is non-self-averaging at long times. 

The paper is organized as follows.  Section~\ref{sec:1} contains the necessary background for the following sections. It presents the model, initial states, quantities, and our previous results about the dynamics and self-averaging properties of the survival probability and inverse participation ratio. In Secs.~\ref{sec:3}, \ref{sec:4}, and \ref{sec:5}, we proceed with the analysis of the distributions of these global quantities. This study is separated by time intervals: short times in  Sec.~\ref{sec:3}, long times in Sec.~\ref{sec:4}, and intermediate times in Sec.~\ref{sec:5}. The analysis of the local quantities and how to use the distribution of one quantity to describe the self-averaging behavior of another one is explained in Sec.~\ref{sec:local}. Conclusions are presented in Sec.~\ref{sec:conclusions}.

\section{Models, Quantities, and Time Scales}
\label{sec:1}

We study two models described by Hamiltonians of the form
\be
H=H_0+ V,
\label{eq:H}
\ee
where $H_0$ is the unperturbed part of the total Hamiltonian and $V$ is a strong perturbation that takes the system into the chaotic regime. The notation adopted is the following: $|n\rangle$ stands for the eigenstates of $H_0$, $|\alpha\rangle$ for the eigenstates of $H$, and $E_\alpha$ for the eigenvalues of $H$. One model consists of random matrices from a GOE and the other is a many-body spin-1/2 system.

\subsection{GOE model}
For the GOE model, $H_0$ is the diagonal part of a full random matrix of dimension $D$ and $V$ contains the off-diagonal elements. The entries are all real random numbers from a Gaussian distribution with mean value $\left<H_{ij}\right>=0$ and variance
\be
\left<H_{ij}^2\right>=
\left\{
\begin{array}{ll}
1 & i=j, \\
1/2 & i\neq j.
\end{array}
\right.
\label{hgoe}
\ee
The Hamiltonian matrix $H$ can be generated by creating a matrix $M$ with random numbers from a Gaussian distribution with mean 0 and variance 1 and then adding $M$ to its transpose as $H=(M+M^{T})/2$~\cite{NoteGOE}. The eigenvalues of this model are highly correlated~\cite{MehtaBook,HaakeBook,Guhr1998} and the eigenstates are normalized random vectors~\cite{Zelevinsky1996}. There are no realistic systems described by this model, but it allows for analytical derivations not only for static properties~\cite{Wigner1958,MehtaBook,Guhr1998}, but also for the dynamics~\cite{Gorin2006,Torres2018,Schiulaz2019,Schiulaz2020}.

\subsection{Disordered spin model}

We consider a one-dimensional chaotic spin-$1/2$ model of great experimental interest~\cite{Schreiber2015} and often used in studies of many-body localization~\cite{SantosEscobar2004,Dukesz2009,Pal2010,Nandkishore2015,Torres2017,Luitz2017}. It has onsite disorder and nearest neighboring couplings~\cite{Avishai2002},
\ba
H_0 &=& J \sum_{k=1}^L (h_k S_k^z +  S_k^z S_{k+1}^z), \nonumber \\
V &=& J\sum_{k=1}^L (S_k^x S_{k+1}^x + S_k^y S_{k+1}^y).
\label{eq:Hspin}
\ea
Above, $\hbar=1$, $J=1$ is the coupling strength, $S_k^{x,y,z}$ are spin operators on site $k$, $L$ is the size of the chain, which is even throughout this work, and periodic boundary conditions are used. The Zeeman splittings $h_i$ are random numbers uniformly distributed in $[-h,h]$. The total magnetization in the $z$ direction is conserved, so we take the largest subspace, where the total $z$ magnetization is zero and the dimension is $D=L!/(L/2)!^2$. We use disorder strength $h=0.75$, which places the system in the chaotic regime. The level statistics and the structure of the eigenstates away from the borders of the spectrum are comparable to those of the GOE model.

\subsection{Initial state}

The initial state $\left| {\rm ini} \right>  = \left| \Psi (0)\right> $ is an eigenstate $|n\rangle $ of $H_0$.
We take $\left|\Psi (0)\right>$ with energy close to the middle of the spectrum, where the eigenstates are chaotic~\cite{TorresKollmar2015},
\be
E_{\rm ini} = \langle \Psi(0)|H|\Psi(0)\rangle = \sum_\alpha\left|C_\alpha^{\rm ini}\right|^2 E_\alpha \sim 0.
\label{Eq:Eini} 
\ee
In the equation above,
\be 
C_\alpha^{\rm ini}=\left<\alpha|\Psi (0)\right>
\ee 
are real components, since the Hamiltonian matrices treated in this work are real and symmetric.
For the spin model, the initial states are product states in the $z$ direction, where on each site the spin either points up or down in the $z$ direction, such as $|\uparrow \downarrow \uparrow \downarrow \downarrow \uparrow \ldots \rangle$. They are often referred to as site-basis vectors or computational basis vectors.

\subsection{Quantities}

We analyze in detail the distributions over disorder realizations of the survival probability and the inverse participation ratio. Both are nonlocal quantities in real space. We also present results for the spin autocorrelation function, its absolute value and its square value, and for the connected spin-spin correlation function. These four quantities are local in space.

Our studies of the survival probability and the inverse participation ratio are presented for the GOE model and the chaotic spin model. For the local quantities, this is done only for the spin model, since the notion of locality does not exist in full random matrices.

The survival probability is the squared overlap of the initial state and its evolved counterpart, 
\ba
P_S(t)&=&\left|\left<\Psi (0)\right|e^{-iHt}\left|\Psi (0)\right>\right|^2 = 
\left| \sum_{\alpha} \left|C_\alpha^{\rm ini}\right|^2 e^{-i E_{\alpha} t} \right|^2 \nonumber \\
&=&\left| \int dE e^{-iEt} \rho_{\rm ini}(E) \right|^2,
\label{eq:PS}
\ea
where
\be
\rho_{\rm ini}(E)=\sum_{\alpha}\left|C_\alpha^{\rm ini}\right|^2\delta(E-E_\alpha)
\label{eq:LDOS}
\ee
is the energy distribution of the initial state. $\rho_{\rm ini}(E)$ is usually referred to as local density of states (LDOS) or strength function. The width $\Gamma$ of this distribution depends on the number of states $|n\rangle $  that are directly coupled with $\left|\Psi (0)\right>$,
\ba
\Gamma^2 
&=& \sum_{\alpha} \left| C_\alpha^{\rm ini} \right|^2 E_{\alpha}^2  - \left( \sum_{\alpha} \left| C_\alpha^{\rm ini} \right|^2 E_{\alpha} \right)^2 \nonumber \\
&=& \langle \Psi(0) |H H |\Psi(0) \rangle - \langle \Psi(0) |H|\Psi(0) \rangle^2 \nonumber \\
&=& \sum_{n } \langle \Psi(0) |H |n\rangle \langle n| H| \Psi(0) \rangle - \langle \Psi(0) |H |\Psi(0) \rangle^2 \nonumber \\
&=&  \sum_{n \neq {\rm ini}} |  \langle n |H| \Psi(0) \rangle |^2.
\label{eq:Gamma}
\ea
The survival probability is a quantity of great theoretical and experimental~\cite{Singh2019} relevance. It has been used in studies of the quantum speed limit~\cite{Bhattacharyya1983,Ufink1993}, onset of exponential~\cite{Jacquod2001,Cerruti2002} and power-law~\cite{Khalfin1958,MugaBook,Campo2016NJP,Tavora2016,Tavora2017} decays, quench dynamics~\cite{Torres2014PRA,Torres2014NJP,Torres2014PRE,Mazza2016,Lerma2018,VolyaARXIV}, ground-state and excited-state quantum phase transitions~\cite{Heyl2013,Santos2016}, quantum scars~\cite{Heller1984,VillasenorARXIV}, multifractality in disordered systems~\cite{Ketzmerick1992,Mirlin2000,Torres2015,Bera2018}, and emergence of the correlation hole~\cite{Leviandier1986,Wilkie1991,Alhassid1992,Gruver1997,Torres2017Philo,Cotler2017,Lerma2019,CruzARXIV,Corps2020}.  

The inverse participation ratio measures the degree of delocalization of a state in a certain basis~\cite{Edwards1972,Izrailev1990,Mirlin2000}. Here, we study a dynamical version of it~\cite{Flambaum2001b,Borgonovi2019R,Borgonovi2019}, which accounts for the spreading in time of the initial many-body state in the basis of unperturbed many-body states  $\left|n\right>$. It is defined as
\be
{\rm{IPR}}(t)=\sum_n\left|\left<n\right|e^{-iHt}\left|\Psi (0)\right>\right|^4 .
\label{eq:IPR}
\ee
At $t=0$, when $|\Psi(0)\rangle$ is one of the states $|n\rangle$, ${\rm{IPR}}(t)=1$. As $|\Psi(0)\rangle$ spreads into other states $|n\rangle$, ${\rm{IPR}}(t)$ decays. For chaotic systems perturbed far from equilibrium, it reaches very small values.

The spin autocorrelation function measures the proximity of a spin $k$ at time $t$ to its orientation at $t=0$ and it is averaged over all sites,
\be
I(t)=\frac{4}{L}\sum_{k=1}^L \left<\Psi (0) \right|S^z_k e^{iHt} S^z_k e^{-iHt}\left|\Psi (0) \right>.
\label{eq:I}
\ee
This quantity is equivalent to the density imbalance between even and odd sites measured in experiments with cold atoms~\cite{Schreiber2015}, as can be seen by mapping the spins into hardcore bosons. The self-averaging behavior of this quantity was studied in Refs.~\cite{Schiulaz2020,Torres2020}. Here, we analyze also $|I(t)|$ and $I^2(t)$. This is done because at long times, $I(t)$ can reach negative values and the oscillations between negative and positive values may complicate the analysis of self-averaging, which is avoided with the other two quantities. 

The connected spin-spin correlation function is given by
\ba
C(t)&=&\frac{4}{L}\sum_k\left[\left<\Psi(t)\right|S_k^zS_{k+1}^z\left|\Psi(t)\right>\right.\\
&-&\left.\left<\Psi(t)\right|S_k^z\left|\Psi(t)\right>\left<\Psi(t)\right|S_{k+1}^z\left|\Psi(t)\right>\right]\nonumber
\ea
and is measured in experiments with ion traps~\cite{Richerme2014}.

\subsection{Self-Averaging and Timescales}
\label{sec:2}

The results presented in this subsection have already appeared in Refs.~\cite{Schiulaz2019,Schiulaz2020}. The purpose of this summary is to serve as a reference for the discussions in the next sections. We show first the evolution of the mean survival probability. The various timescales involved in the relaxation process of this quantity are the ones used in the analysis of the distributions of all quantities in the next sections. We also describe here the time-dependence of the relative variance of the survival probability and of the inverse participation ratio, whose distributions are the subjects of Secs.~\ref{sec:3}, \ref{sec:4}, and ~\ref{sec:5}.

A quantity $O$ is self-averaging when its relative variance
\be
{\cal R}_{O}(t)= \frac{\sigma_{O}^2 (t) }{\left<O(t)\right>^2} = \frac{\left<O^2(t)\right>-\left<O(t)\right>^2}{\left<O(t)\right>^2} 
\label{eq:sigma} 
\ee
decreases as the system size increases. The notation $\left< \cdot \right>$ indicates in our case the average over disorder realizations and also initial states. We consider $0.01D$ initial states and at least $10^4/(0.01D)$ disorder realizations, so that each point for the curves of $\left<O(t)\right>$ and ${\cal R}_{O}(t)$ is an average over $10^4$ data.

\subsubsection{Survival probability}

\begin{figure}[h!]
\includegraphics*[width=0.46\textwidth]{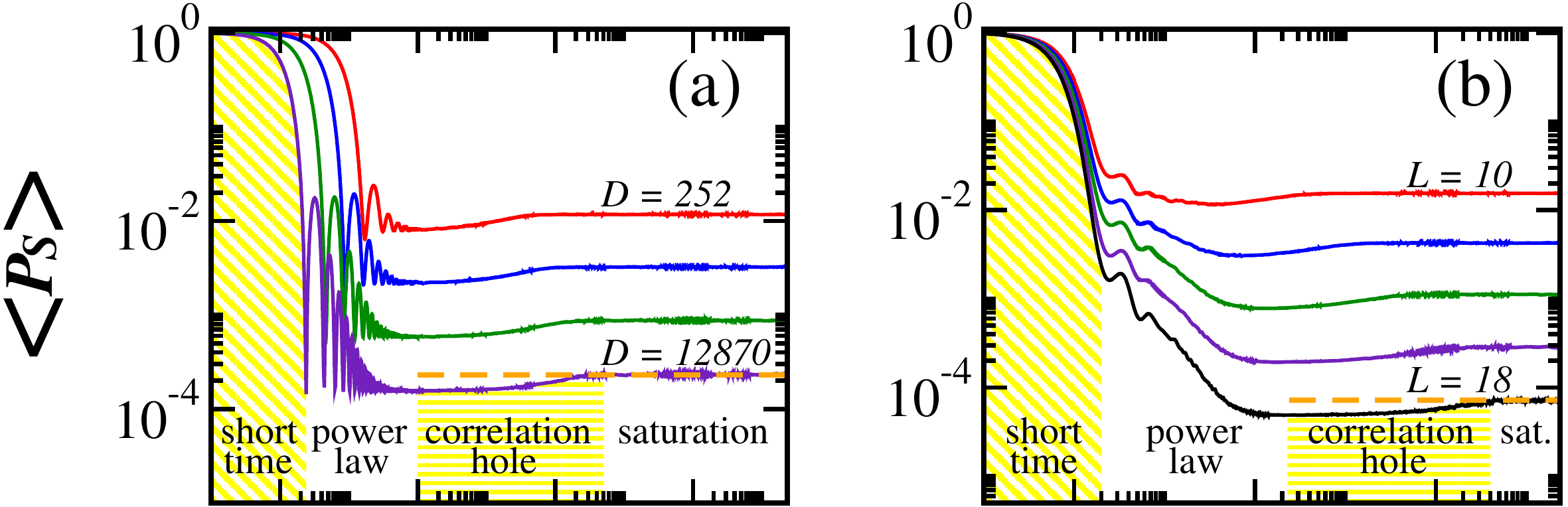}\\
\vspace{0.2cm}
\includegraphics*[width=0.46\textwidth]{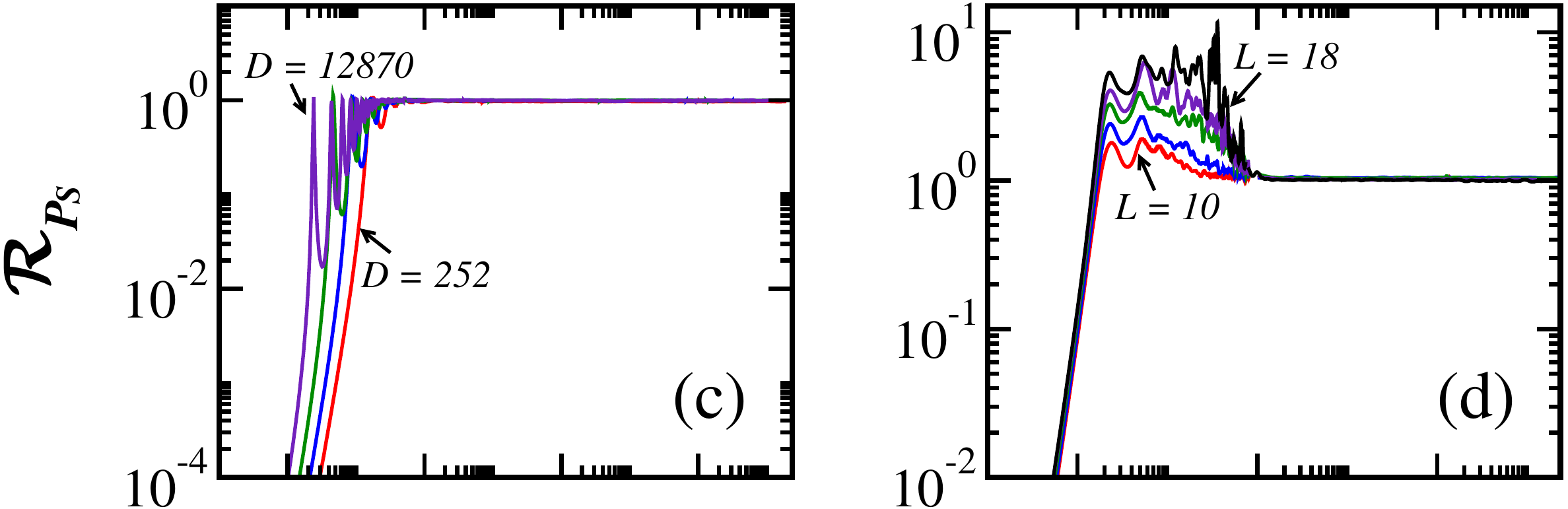}\\
\includegraphics*[width=0.46\textwidth]{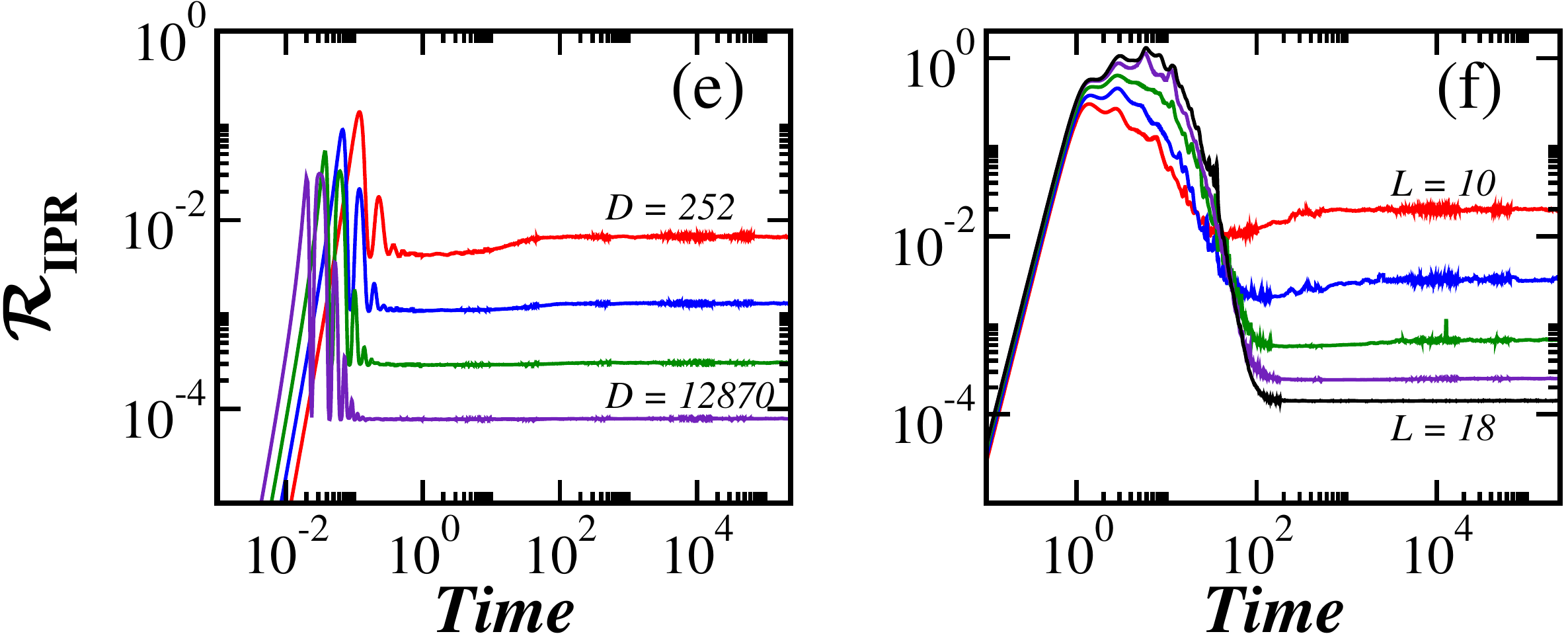}
\caption{Evolution of the mean of the survival probability (a, b), of the relative variance of the survival probability (c, d), and of the relative variance of the inverse participation ratio (e, f) for the GOE model (left panels) and the chaotic disordered spin model (right panels). The time intervals for the fast initial decay, power-law behavior, correlation hole, and saturation are indicated in panels (a) and (b). The horizontal dashed line marks the saturation value of $P_S$ for the largest size. System sizes: $D=252, 924, 3\,432, 12\,870$ ($L=10,12, 14, 16$). For the spin model, $L=18$ is also shown.  In all panels: $0.01D$ disorder realizations and $10^4/(0.01D)$ initial states.}
\label{fig01}       
\end{figure}

The top panels of Fig.~\ref{fig01} show the survival probability for the GOE model [Fig.~\ref{fig01}~(a)] and the spin model [Fig.~\ref{fig01}~(b)]. The shape and bounds of the LDOS [Eq.~(\ref{eq:LDOS})] determine the initial decay of the survival probability. The LDOS for the GOE model is semicircular. The square of the Fourier transform of a semicircle gives $\mathcal{J}^2_1(2 \Gamma t)/(\Gamma^2 t^2)$, where ${\cal J}_1$ indicates the Bessel function of the first kind~\cite{Wigner1955}. This implies that after a very rapid initial decay, $\langle P_S(t) \rangle $ shows oscillations that decay according to a power law $\propto t^{-3}$ \cite{Tavora2016,Tavora2017,Cotler2017}, as seen in Fig.~\ref{fig01}~(a). The LDOS for the spin model is Gaussian~\cite{Torres2014PRA,Torres2014NJP}, as found in many-body quantum systems with two-body couplings and perturbed far from equilibrium~\cite{Zelevinsky1996,Frazier1996,Flambaum2001a,Flambaum2001b,Angom2004,Kota2006}. The square of the Fourier transform of a bounded Gaussian gives $\exp(-\Gamma^2 t^2) {\cal F}(t)/(4 {\cal N}^2)$, where ${\cal F}(t)$ involves error functions and ${\cal N}$ is a normalization constant (see the appendices in Refs.~\cite{Tavora2017,Schiulaz2019}). This implies that after an initial Gaussian decay~\cite{Torres2014PRA,Torres2014NJP}, $\langle P_S(t) \rangle $ shows a power-law behavior $\propto t^{-2}$ \cite{Tavora2016,Tavora2017,Torres2019EPJ}, as observed in Fig.~\ref{fig01}~(b). The origin of the power-law decay of the survival probability in bounded spectra has been discussed at least since the 1950's ~\cite{Khalfin1958,Erdelyi1956,Ersak1969,Fleming1973,Fonda1978,Urbanowski2009} and more recently in Ref.~\cite{Yang2020}.  The experimental detection of algebraic decay at long times has been reported in Ref.~\cite{Rothe2006}, and evidence of slower relaxation for the density imbalance in the context of many-body localization of one- and two-dimensional quasiperiodic systems was presented in Refs.~\cite{Lschen2017,Bordia2017}. 

The power-law decays in Figs.~\ref{fig01}~(a) and~\ref{fig01}~(b) persist up to a time denoted by $t_{\rm{Th}}$~\cite{Schiulaz2019}, where $\langle P_S(t) \rangle $ reaches its minimum value. Beyond this point, the survival probability increases until the dynamics saturates for $t>t_{\rm{R}}$, where $t_{\rm{R}}$ is the relaxation time. At this point, $\langle P_S (t>t_{\rm{R}})\rangle$ fluctuates around the infinite-time average $ \langle \sum_{\alpha} \left|C_\alpha^{\rm ini}\right|^4 \rangle$. The dip below the saturation point is known as correlation hole~\cite{Leviandier1986,Wilkie1991,Alhassid1992} and it appears only in systems where the eigenvalues are correlated, reflecting short- and long-range correlations~\cite{Ma1995}.

The four time intervals for the distinct behaviors of $\langle P_S(t) \rangle $ -- fast initial decay, power-law behavior, correlation hole, and saturation -- are indicated in Figs.~\ref{fig01}~(a) and~\ref{fig01}~(b). These are the timescales that we consider in the next sections to investigate the distributions of the survival probability and of the other quantities as well. 

In Figs.~\ref{fig01}~(c) and~\ref{fig01}~(d), we show the results for the relative variance ${\cal R}_{P_S}(t)$
for different system sizes. The survival probability is non-self-averaging at any timescale, as shown analytically in Ref.~\cite{Schiulaz2020}. Initially, ${\cal R}_{P_S}(t)$ grows with system size, while for $t>t_{\rm{Th}}$, it reaches a constant value, ${\cal R}_{P_S}(t) \sim 1$. There is no noticeable difference between the value of ${\cal R}_{P_S}(t) $ in the interval $[t_{\rm{Th}}, t_{\rm{R}}]$ and for $t>t_{\rm{R}}$.

\subsubsection{Inverse participation ratio}

Plots for the mean of the inverse participation ratio can be seen in Ref.~\cite{Schiulaz2020}. There are two different behaviors for $\langle {\rm{IPR}}(t) \rangle$ at short times. The decay is initially very fast and then it either oscillates in the case of the GOE model or it slows down for the spin model. These two timescales coincide with the intervals for the fast decay and the power-law behavior of $\langle P_S(t) \rangle$. Beyond this point, however, a correlation hole is not visible for $\langle {\rm{IPR}}(t) \rangle$. It exists, but it is extremely small~\cite{Schiulaz2019} and, contrary to what we find for the survival probability, the ratio between the saturation point of $\langle {\rm{IPR}}(t) \rangle$ and  its minimum value at the correlation hole decreases as the system size increases.

The evolution of the relative variance of $\rm{IPR}$  is seen in Figs.~\ref{fig01}~(e) and~\ref{fig01}~(f). It shows that the inverse participation ratio is non-self-averaging at short times, which is understandable, since for small times,$\langle {\rm{IPR}}(t) \rangle \sim \langle P_S^2(t) \rangle$.  But for times $t>t_{\rm Th}$, the inverse participation ratio becomes ``super'' self-averaging, by which we mean that ${\cal R}_{\rm IPR}(t) \propto 1/D$  instead of  $\propto 1/L$.

\section{Distributions at short times}
\label{sec:3}

\begin{figure}[ht]
\includegraphics*[width=0.45\textwidth]{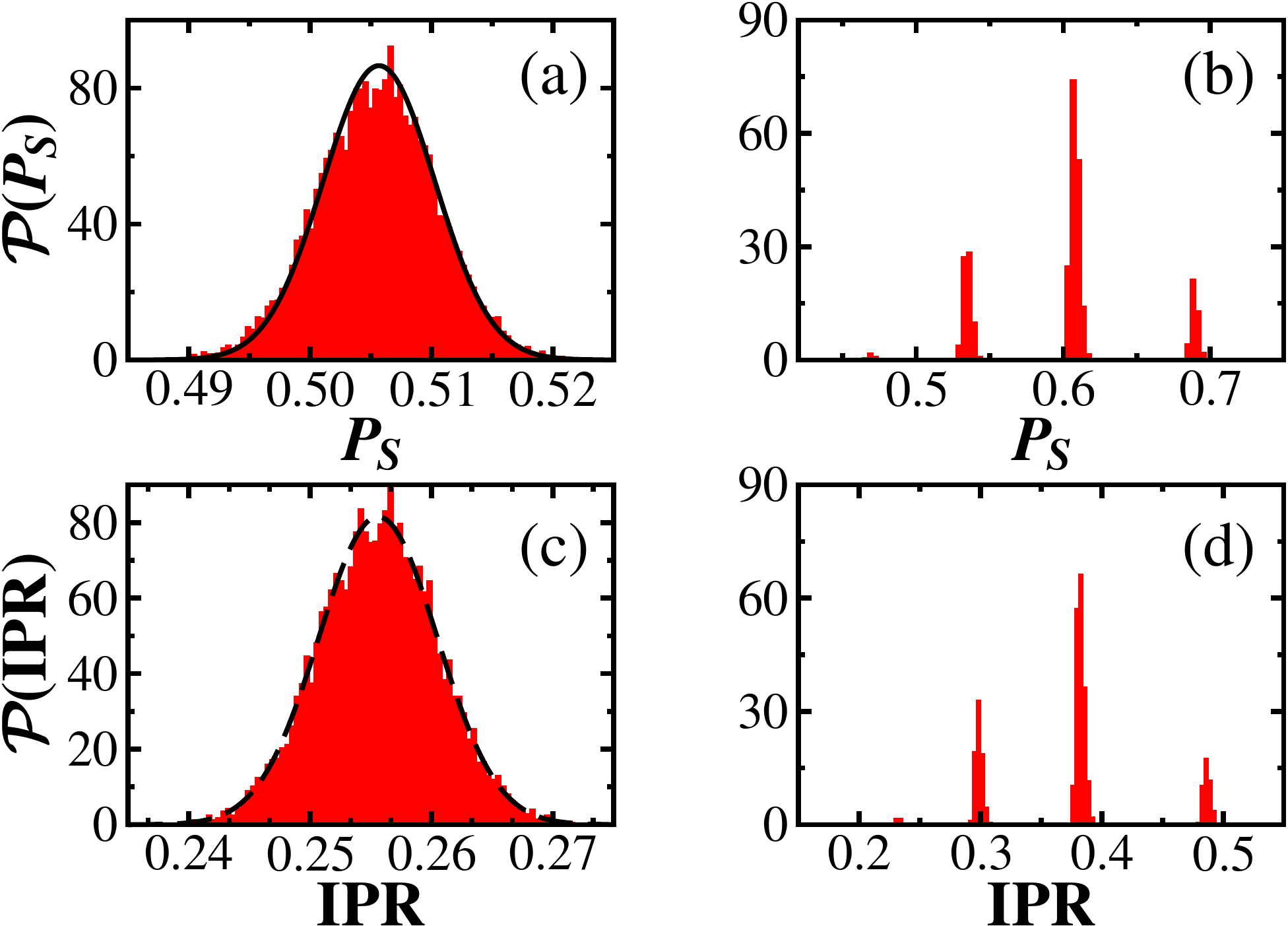}
\caption{Distributions of the survival probability (a, b) and inverse participation ratio (c, d) for the GOE  (a, c) and the spin (b, d) model at very short times: $t=0.01$ and $t=0.5$, respectively. Solid line in (a) is the theoretical Gaussian distribution with mean from Eq.~(\ref{EqSPmean}) and variance  from Eq.~(\ref{EqSPvar}), and dashed line in (c) is the Gaussian with the numerical values for the mean and variance.}
\label{fig:short}       
\end{figure}

In Fig.~\ref{fig:short}, we show the distributions of the survival probability [Figs.~\ref{fig:short}~(a) and~\ref{fig:short}~(b)] and of the inverse participation ratio [Figs.~\ref{fig:short}~(c) and~\ref{fig:short}~(d)] for the GOE model [Figs.~\ref{fig:short}~(a) and~\ref{fig:short}~(c)] and the spin model [Figs.~\ref{fig:short}~(b) and~\ref{fig:short}~(d)] at short times, $t < \Gamma^{-1}$, when the decays of $\langle P_S(t) \rangle$ and $\langle {\rm IPR}(t) \rangle$ are very fast. The distributions are similar for both quantities, but differ between the models. 

At short times, the main contribution for $\langle {\rm IPR}(t) \rangle$ is the square of the survival probability, $\langle {\rm IPR}(t\ll \Gamma^{-1}) \rangle \sim \left|\left< \Psi (0) \right| e^{-iHt}\left|\Psi (0)\right>\right|^4$, which explains why the distributions for both quantities are so similar. Compare Fig.~\ref{fig:short}~(a) with Fig.~\ref{fig:short}~(c), and Fig.~\ref{fig:short}~(b)  with Fig.~\ref{fig:short}~(d). Therefore, it suffices to describe below the distributions for the survival probability.

\subsection{Survival probability}
At short times, the decay of the survival probability is controlled by the short-time expansion of $\mathcal{J}^2_1(2 \Gamma t)/(\Gamma^2 t^2)$ for the GOE model and of $\exp(-\Gamma^2 t^2)$ for the spin model. The distribution of $P_S(t)$ at a fixed time $t< \Gamma^{-1}$ reflects then the distribution of the square of the width of the LDOS, $\Gamma^2$, and its higher powers. 

\subsubsection{Survival probability: GOE model}
For the GOE model, the expansion gives

\[\frac{\mathcal{J}^2_1(2 \Gamma t)}{\Gamma^2 t^2}= 1 - \Gamma^2 t^2 + \frac{5}{12}\Gamma^4 t^4  - \frac{7}{72}\Gamma^6 t^6 + \frac{7}{480} \Gamma^8 t^8  \dots
\]

As we saw in Eq.~(\ref{eq:Gamma}), $\Gamma^2$ is the sum of the square of the off-diagonal elements contained in the row of the Hamiltonian matrix where the initial state lies. For the GOE model, this means the sum of the square of $D-1$ Gaussian random numbers with  $\langle H_{ij}\rangle =0$ and  $\langle H_{ij}^2\rangle=1/2$, which gives a $\chi^2$-distribution with $D-1$ degrees of freedom. This is approximately a Gaussian distribution with mean $\mu_{\Gamma^2 }=(D-1)/2$ and variance $\sigma_{\Gamma^2 }^2 =(D-1)/2$. 

Using $g_n$ as a notation for the moments of $\Gamma^2$,  that is, 
\[
g_n= \frac{1}{ \sqrt{2 \pi \sigma_{\Gamma^2 }^2}}\int (\Gamma^2)^n \exp\left[- \frac{(\Gamma^2 -\mu_{\Gamma^2 })^2}{2  \sigma_{\Gamma^2 }^2}\right]d\Gamma^2,
\]
and keeping terms up to 8th order in time we see that
\be
\langle P_S (t) \rangle \approx 1 - g_1 t^2 + \frac{5}{12}g_2 t^4 - \frac{7}{72}g_3 t^6 + \frac{7}{480}g_4 t^8,
\label{EqSPmean}
\ee
and the variance  
\ba
\label{EqSPvar}
&&  \sigma_{P_S(t)}^2     \\
&&  =(g_2 - g_1^2) t^4 
- \frac{5}{6} (g_3 - g_1 g_2) t^6 \nonumber \\
&&  + \left[ \frac{25}{144} (g_4 - g_2^2) +\frac{7}{36} (g_4 - g_1 g_3) \right]t^8 \nonumber \\
&&  - \left[ \frac{7}{240} (g_5 - g_1 g_4)  + \frac{35}{432} (g_5 - g_2 g_3) \right] t^{10} 
\nonumber \\
&&  +
\left[     \frac{49}{5184} (g_6 \!-\! g_3^2) +  \frac{11}{3600}  (g_6 \!-\! g_1 g_5) + \frac{7}{576} (g_6 \!-\! g_2 g_4)
\right] t^{12} . \nonumber
\ea
For a fixed $t=0.01 $ and $D=12\,870$, $\langle P_S (0.01) \rangle \sim 0.505$ and $\sigma_{P_S(0.01)}^2 \sim 2.1 \times 10^{-5} $, which are the values used in the Gaussian indicated with a solid line in Fig.~\ref{fig:short}~(a).

\subsubsection{Survival probability: Spin model}
For the spin model, the energy $E_{\text{ini}}$ [Eq.~(\ref{Eq:Eini})] of the initial state depends on the disorder strength and on the number $n_p$ of neighboring pairs of up-spins as determined by the Ising interaction, $\sum_k S_k^z S_{k+1}^z$. Focusing only on the Ising interaction, one can see that it leads to $L/2$ energy bands that go from the band of lowest energy with no pairs of up-spins, which has only the two N\'eel states  $|\uparrow \downarrow \uparrow \downarrow \ldots \rangle$ and $|\downarrow \uparrow \downarrow \uparrow \ldots \rangle$, to the band of highest energy with $n_p=L/2-1$ neighboring pairs of up-spins, which has $L$ states~\cite{Joel2013}. The number of states in a band grows as we approach the middle of the spectrum. The most populated band for chain sizes that are multiple of 4 is centered at energy zero, and for the chains of other even sizes, it is centered at $-1/2$.

The fragmented distribution in Fig.~\ref{fig:short}~(b) reflects the bands created by the Ising interaction. Each state in a band with $n_p$ pairs of neighboring up-spins couples with $(L-2n_p)$ other states, so according to Eq.~(\ref{eq:Gamma}), $\Gamma^2=(L-2n_p)/4$. For the $L=16$ case shown in Fig.~\ref{fig:short}~(b), the states in the most populated band at energy zero has $n_p=4$ and $\Gamma^2=2$, so $P_S(t<\Gamma^{-1}) \sim \exp(-\Gamma^2 t^2)$ gives $\sim 0.61$ for $t=0.5$, which is indeed the center of the highest peak in Fig.~\ref{fig:short}~(b). The two other highest peaks correspond to the Ising band at $-1$ with $n_p=3$ and $P_S(0.5)\sim 0.54$ and the band at $1$ with $n_p=5$ and $P_S(0.5)  \sim 0.69$.

As discussed in Ref.~\cite{Schiulaz2020}, both the survival probability and in the inverse participation ratio are non-self-averaging at short times. This can be understood from the expansion of the survival probability at the lowest order in $t$,
\be
P_S(t) \sim 1 - \Gamma^2 t^2,
\ee
which gives
\ba
{\cal R}_{P_S}(t) &\sim&\frac{\left<  (1 - \Gamma^2 t^2)^2 \right>-\left<1 - \Gamma^2 t^2\right>^2}{\left<1 - \Gamma^2 t^2\right>^2}  \nonumber \\
&=& \sigma_{\Gamma^2 }^2 t^4 .
\ea
The lack of self-averaging happens because $\sigma_{\Gamma^2 }^2 $ grows with $L$ for the spin model and with $D$ for the GOE model, having no relationship with the shape of the distributions.

\section{Distributions after Saturation}
\label{sec:4}

In Fig.~\ref{fig:long}, we show the distributions of $P_S(t)$ and ${\rm IPR}(t)$ after the saturation of the dynamics, for a fixed time $t>t_{\rm R}$. In contrast with the behavior at short times, the distributions for both models are now similar, while they differ between quantities. In realistic chaotic systems, properties similar to those of random matrices manifest themselves at long times.

\subsection{Survival probability}

\begin{figure}[htb]
\includegraphics[width=0.47\textwidth]{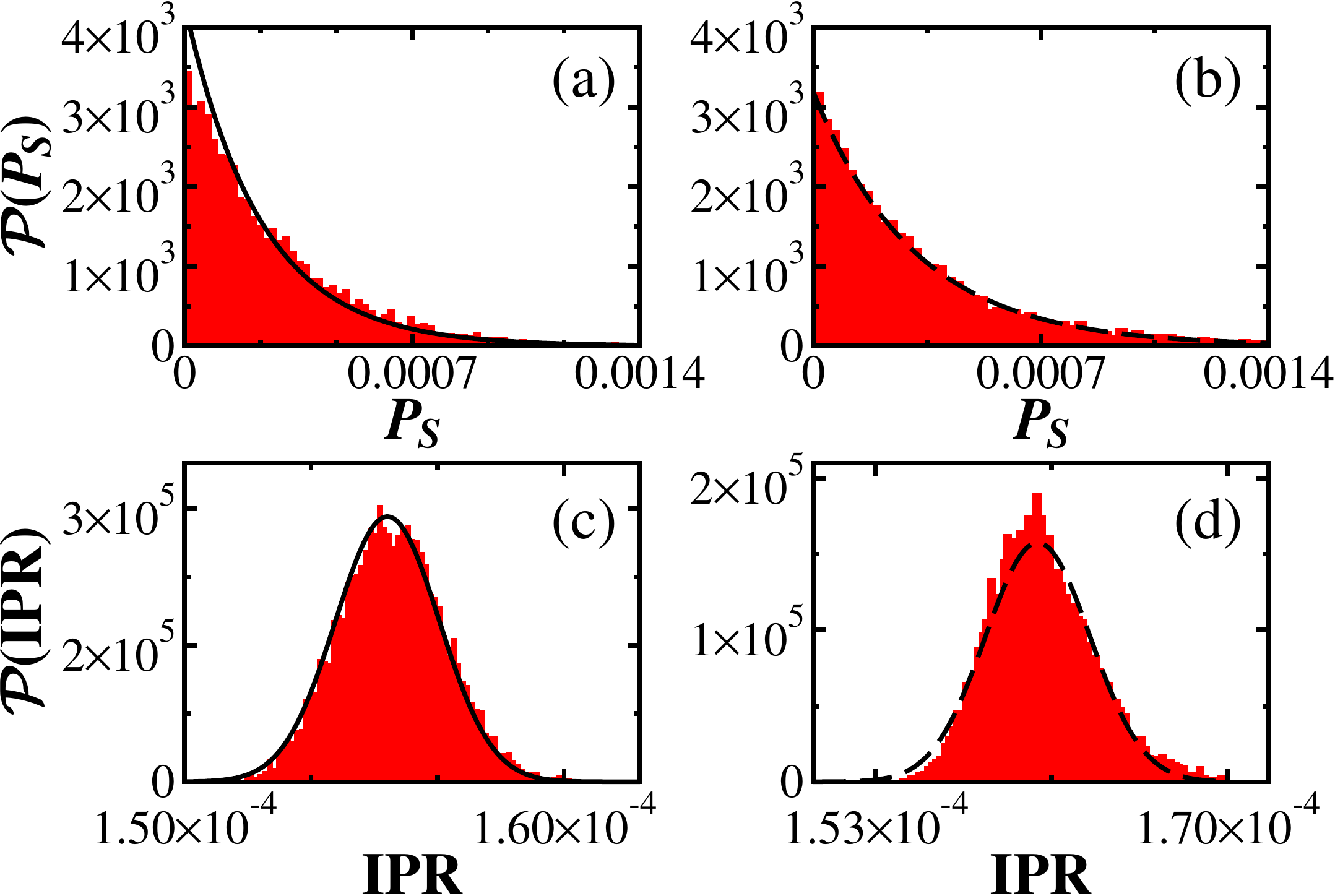}
\caption{Distributions of the survival probability (a, b) and inverse participation ratio (c, d) for the GOE model  (a, c) at $t=10^3$  and for the spin model (b, d) at $t=5\times10^4 $. Solid lines in (a) are the exponential distribution with rate parameter $D/3$, and in (c) they are the Gaussian distribution with mean and variance from Eqs.~(\ref{IPRav}) and~(\ref{IPRsig}). Dashed lines in (b) are the exponential distribution with rate parameter $1/\langle \sum_{\alpha } |C_{\alpha}^{\rm ini}|^4 \rangle$, and in (d) they are the Gaussian curve with the numerical values for $\langle \text{IPR} (t) \rangle$ and $\sigma_{\text{IPR}} (t)$ at $t=5\times10^4$.}
\label{fig:long}       
\end{figure}

The distribution of $P_S(t)$ for the GOE and the spin model for $t>t_{\rm R}$ is exponential, as shown in Figs.~\ref{fig:long}~(a) and~\ref{fig:long}~(b).  Since the mean and the dispersion of exponential distributions are equal, ${\cal R}_{P_S}(t>t_{\rm R}) \sim 1$, as indeed found numerically in Figs.~\ref{fig01}~(c) and~\ref{fig01}~(d). This justifies the lack of self-averaging of the survival probability for $t>t_{\rm R}$.

The rate parameter of an exponential distribution is the reciprocal of the mean. For the distribution of $P_S(t>t_{\rm R})$, the rate parameter is  $1/\sum_{\alpha } |C_{\alpha}^{\rm ini}|^4$. This can be understood by writing the survival probability as
\be
P_S (t) =  \sum_{\alpha < \beta}  2 |C_{\alpha}^{\rm ini}|^2 |C_{\beta}^{\rm ini}|^2 \cos \left[ (E_{\alpha} - E_{\beta}) t \right] + \sum_{\alpha } |C_{\alpha}^{\rm ini}|^4 .
\label{SPcos}
\ee
On average, the first term on the right hand side of the equation above cancels out, so $
\langle P_S(t>t_{\text{R}}) \rangle \sim  \sum_{\alpha } |C_{\alpha}^{\rm ini}|^4$. 

The eigenstates of the GOE model are random vectors, so  $C_{\alpha}^{\rm ini}$'s are random numbers from a Gaussian distribution satisfying the constraint $\sum_{\alpha=1}^D |C_{\alpha}^{\rm ini}|^2 =1$. Using ${\cal P}(C) = \sqrt{D/(2\pi)} e^{-DC^2/2}$ for the components, we have $\langle C \rangle $=0,  $\langle C^2 \rangle=1/D$, and $\langle C^4 \rangle=3/D^2$, so $\sum_{\alpha } |C_{\alpha}^{\rm ini}|^4=\sum_{\alpha } (3/D^2) = 3/D$. 

For the chaotic spin model, the eigenstates away from the edges of the spectrum are also approximately random vectors, so  $\sum_{\alpha } |C_{\alpha}^{\rm ini}|^4$ is close to $3/D$, although slightly larger. This discrepancy becomes particularly evident if one fits the numerical distribution in Fig.~\ref{fig:long}~(b) with a single parameter.  The fact that we get a value slightly larger than $3/D$ indicates some remaining degree of correlations between the components of the initial state.

A simple justification for the exponential shape of the distribution for $P_S(t)$ can be given by substituting 
\[
\left| \sum_{\alpha} \left|C_\alpha^{\rm ini}\right|^2 e^{-i E_{\alpha} t} \right|^2,
\]
with
\be
 \frac{1}{D^2} \left\{
\left[ \sum_{\alpha} \cos(E_{\alpha} t) \right]^2 + \left[ \sum_{\alpha} \sin(E_{\alpha} t) \right]^2 
\right\} .
\label{eq:help}
\ee
The sum of the cosines and the sum of the sines are Gaussian random variables, as discussed in Ref.~\cite{Kunz1999} for full random matrices.  The distribution of the sum of the square of two Gaussian random numbers is exponential, which explains the shape seen in Figs.~\ref{fig:long}~(a) and~\ref{fig:long}~(b). Notice, however, that this simplification gives $1/D$ as the mean value for $P_S(t)$, which differs from the correct value by a factor of 3. Furthermore, we verified numerically that for $t>t_\text{R}$, the sum of the cosines and the sum of the sines are Gaussian random variables also when $E_{\alpha}$ are random numbers from a Gaussian distribution, which indicates that for such long times, the correlations between the eigenvalues are not essential for the onset of the exponential shape of the distribution for $P_S(t)$. This means that even for an integrable model with uncorrelated eigenvalues, the distribution of $P_S(t>t_{\rm R})$ is exponential.

The proper derivation of the exponential distribution for $P_S (t)$ involves the convolution of the distribution for the components of the initial state with the distribution for $e^{-i E_{\alpha} t}$, as done in Ref.~\cite{Kunz2002} for random matrices. The result for $t>t_\text{R}$ is 
\be
{\cal P}(P_S) =\frac{1}{\sum_{\alpha } |C_{\alpha}^{\rm ini}|^4} \exp \left[- \frac{P_S(t) }{\sum_{\alpha } |C_{\alpha}^{\rm ini}|^4} \right].
\ee
The agreement between this theoretical curve and the numerical distribution of $P_S (t)$ for the GOE and also for the spin model is excellent, as seen in Figs.~\ref{fig:long}~(a) and~\ref{fig:long}~(b). 

\subsection{Inverse participation ratio}

The distribution of the inverse participation ratio for the GOE and the spin model at a fixed time $t>t_{\text{R}}$ is Gaussian, as evident in Figs.~\ref{fig:long}~(c) and~\ref{fig:long}~(d). Following the steps described in Ref.~\cite{Kunz2002}, it should be possible to formally derive the Gaussian distribution by doing the convolutions between the distributions for the components $C_{\alpha}^n$ and $C_{\alpha}^{\rm ini} $, which are nearly Gaussian random numbers, and for $e^{- E_{\alpha} t}$. Taking into account the sum over all basis vectors $|n\rangle$ in
\be
 {\rm IPR}(t) = \sum_n \left| \sum_{\alpha} C_{\alpha}^n C_{\alpha}^{\rm ini} e^{- E_{\alpha} t}\right|^4,
 \ee
which is a large sum, one should arrive at the Gaussian shape.

The mean of the distribution of $ {\rm IPR}(t)$ is obtained by realizing that the only terms in
\ba
&& {\rm IPR}(t)  \nonumber \\
&& \!\! = \sum_n \!\!\! \sum_{\alpha, \beta, \gamma, \delta} \!\!\! C_{\alpha}^n C_{\alpha}^{\rm ini}
C_{\beta}^n C_{\beta}^{\rm ini}
C_{\gamma}^n C_{\gamma}^{\rm ini}
C_{\delta}^n C_{\delta}^{\rm ini} e^{- (E_{\alpha} - E_{\beta} + E_{\gamma} - E_{\delta})t}
\nonumber
\ea
that do not average out at long times are those where
$\alpha = \beta$, $\gamma = \delta$, with $\alpha \neq \delta$; $\alpha = \delta$, $\beta=\gamma $, with $\alpha \neq \beta$; and $\alpha = \beta=\gamma = \delta$, which gives
\[
 2 \sum_{n}   \left(    \sum_{\alpha } |C_{\alpha}^{n}|^2 |C_{\alpha}^{\rm ini}|^2    \right)^2 - 
\sum_{\alpha } |C_{\alpha}^{\rm ini}|^4 \left( \sum_{n} |C_{\alpha}^{n}|^4 \right).
\]
Since for the random matrices, $|C_{\alpha}^{n}|^2 \sim 1/D$, we have 
\be
\frac{2}{D} - \frac{9}{D^2}.
\label{IPRav}
\ee

To compute the variance of the distribution, we need the dominant terms of 
\ba
&&  {\rm IPR}^2 (t) \nonumber \\
&&  =\sum_{n} \sum_{\alpha, \beta, \gamma, \delta} \sum_{n'}  \sum_{\alpha', \beta', \gamma', \delta'} \nonumber \\
&& \times
 C_{\alpha}^n C_{\alpha}^{\rm ini}
C_{\beta}^n C_{\beta}^{\rm ini}
C_{\gamma}^n C_{\gamma}^{\rm ini}
C_{\delta}^n C_{\delta}^{\rm ini}
e^{- (E_{\alpha} - E_{\beta} + E_{\gamma} - E_{\delta})t} \nonumber \\
&& \times
C_{\alpha'}^{n'}  C_{\alpha'}^{\rm ini}
C_{\beta'}^{n'} C_{\beta'}^{\rm ini}
C_{\gamma'}^{n'}  C_{\gamma'}^{\rm ini}
C_{\delta'}^{n'}  C_{\delta'}^{\rm ini}
e^{- (E_{\alpha'} - E_{\beta'} + E_{\gamma'} - E_{\delta'})t} .  \nonumber
\ea
There are four terms similar to the one with $\alpha=\beta$, $\alpha'=\beta'$, $\gamma = \delta$, $\gamma' = \delta'$,  which gives 
$4 \sum_{n}   \left(    \sum_{\alpha } |C_{\alpha}^{n}|^2 |C_{\alpha}^{\rm ini}|^2    \right)^2 -  4 \sum_{\alpha } |C_{\alpha}^{\rm ini}|^4 \left( \sum_{n} |C_{\alpha}^{n}|^4   \right) $, and they cancel the dominant terms of $\langle {\rm IPR} (t>t_{\rm R}) \rangle^2$, so they do not contribute to the variance. But there are also four terms similar to the one with $\alpha=\delta$, $\alpha'=\delta'$, $\beta=\gamma $, $\beta'=\gamma' $, which for $n=n'$ gives
\[
4 \sum_{n} \!\! \sum_{\alpha, \beta, \gamma, \delta} \!\!  |C_{\alpha}^n|^2  |C_{\alpha}^{\rm ini} |^2 
| C_{\beta}^n |^2  |C_{\beta}^{\rm ini} |^2 
| C_{\gamma}^n |^2  |C_{\gamma}^{\rm ini} |^2 
| C_{\delta}^n |^2  |C_{\delta}^{\rm ini}|^2,
\]
so the variance of the distribution of ${\rm IPR} (t)$ for a fixed $t>t_{\rm R}$ is
\be
\sigma^2_{\rm IPR}  \sim \frac{4}{D^3}.
\label{IPRsig}
\ee

The Gaussian distribution with the mean from Eq.~(\ref{IPRav}) and the variance from Eq.~(\ref{IPRsig}) matches very well the histogram for the GOE model in Fig.~\ref{fig:long}~(c). Furthermore, our numerical analysis of the distributions obtained for random matrices of different sizes shows that the skewness $\rightarrow 0$ and the kurtosis $\rightarrow 3$ as the dimension of the matrices increases, just as we would expect for a symmetric Gaussian distribution. 

For the spin model, the dashed line in Fig.~\ref{fig:long}~(d) is a Gaussian curve with the numerical values obtained for $\langle \text{IPR} (t) \rangle$ and $\sigma^2_{\text{IPR}} (t)$  for a fixed $t>t_{\text{R}}$. The mean and variance for this curve are slightly larger than the values in Eqs.~(\ref{IPRav}) and~(\ref{IPRsig}), indicating again some degree of correlation between the components of the eigenstates of the realistic model. We might expect the results to approach those for the GOE model as $L$ increases, although our numerical analysis of the distributions for $L=10, 12, 14, 16, 18$ indicates that the skewness $\rightarrow 1$ and the kurtosis $\rightarrow 4$ as the system size increases. These values indicate a nonsymmetric distribution with heavier tails than a Gaussian distribution. 

The results for the mean and variance of ${\rm IPR}(t)$ in Eqs.~(\ref{IPRav}) and~(\ref{IPRsig}) make it clear that ${\cal R}_{\rm IPR}(t)$ decreases as $1/D$ and therefore, the inverse participation ratio becomes self-averaging at long times. The dependence of ${\cal R}_{\rm IPR}(t>t_{\rm R})$ on the dimension of the Hamiltonian matrix instead of the system size $L$ is characteristic of interacting many-body quantum systems. This is related to the fact that the spread of the initial state takes place in the many-body Hilbert space instead of the real space.

\section{Distributions at intermediate times}
\label{sec:5}

As time grows from zero, the distributions for the various quantities studied here gradually change their shapes from those observed at short times (Fig.~\ref{fig:short}) to those at long times (Fig.~\ref{fig:long}). Illustrations of the distributions for $P_S(t)$ and $\text{IPR}(t)$ for the spin model at intermediate times are shown in Fig.~\ref{fig:interm} and discussed below.

\subsection{Survival probability}

\begin{figure}[ht]
\includegraphics*[width=0.4\textwidth]{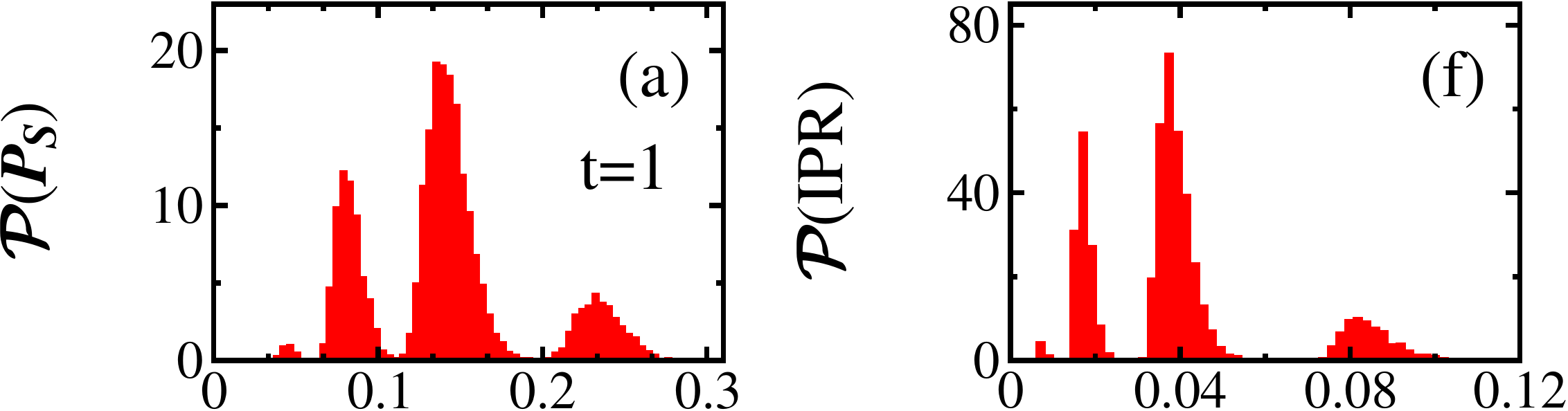}\\
\vspace{0.1cm}
\includegraphics*[width=0.4\textwidth]{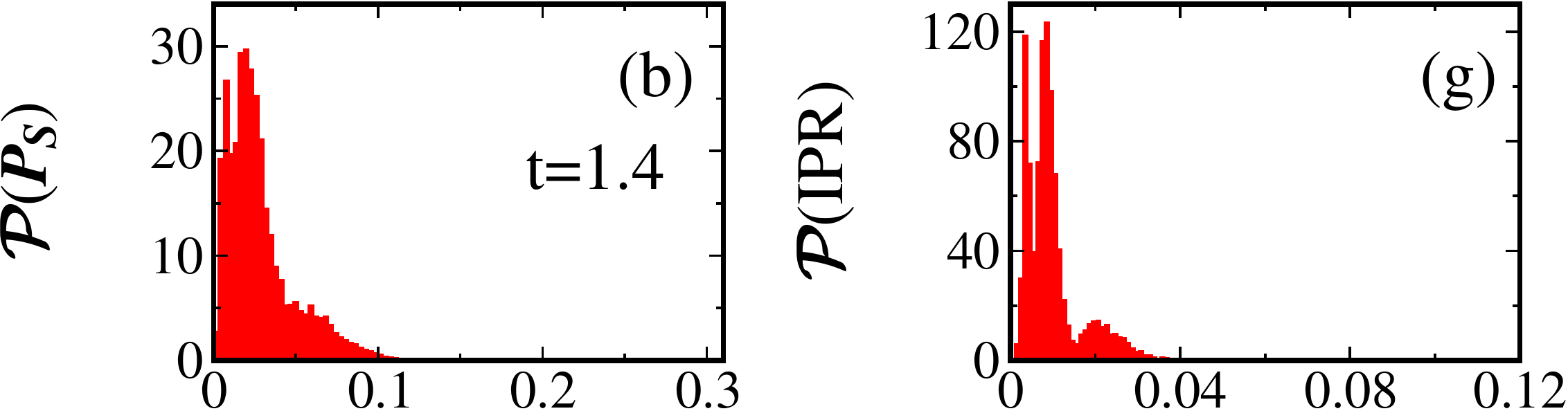}\\
\includegraphics*[width=0.4\textwidth]{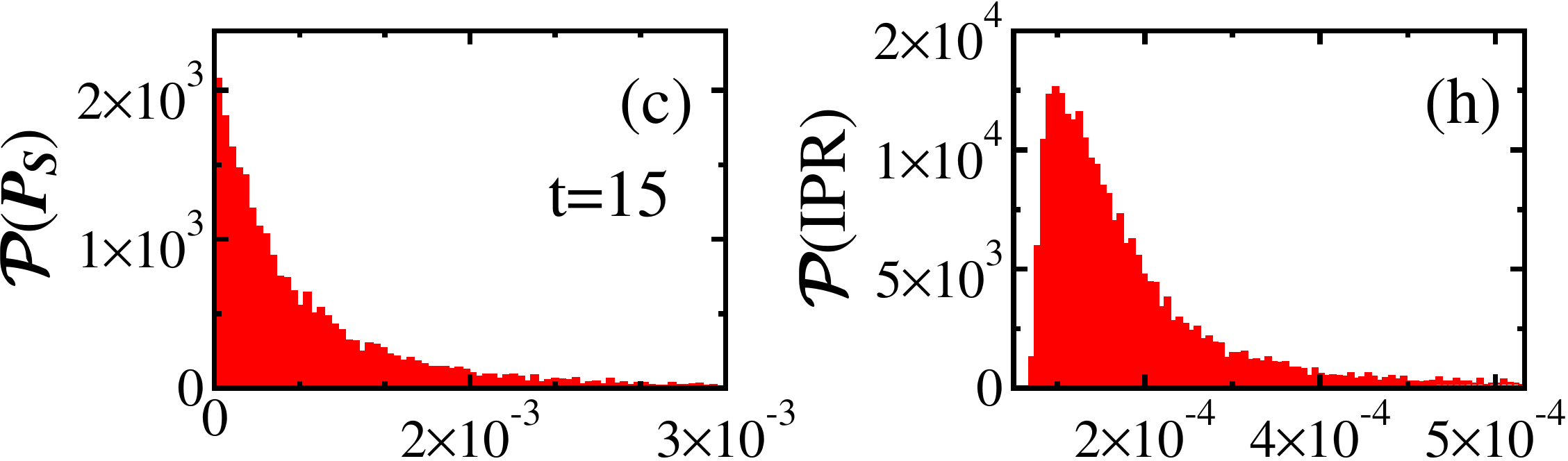}\\
\includegraphics*[width=0.41\textwidth]{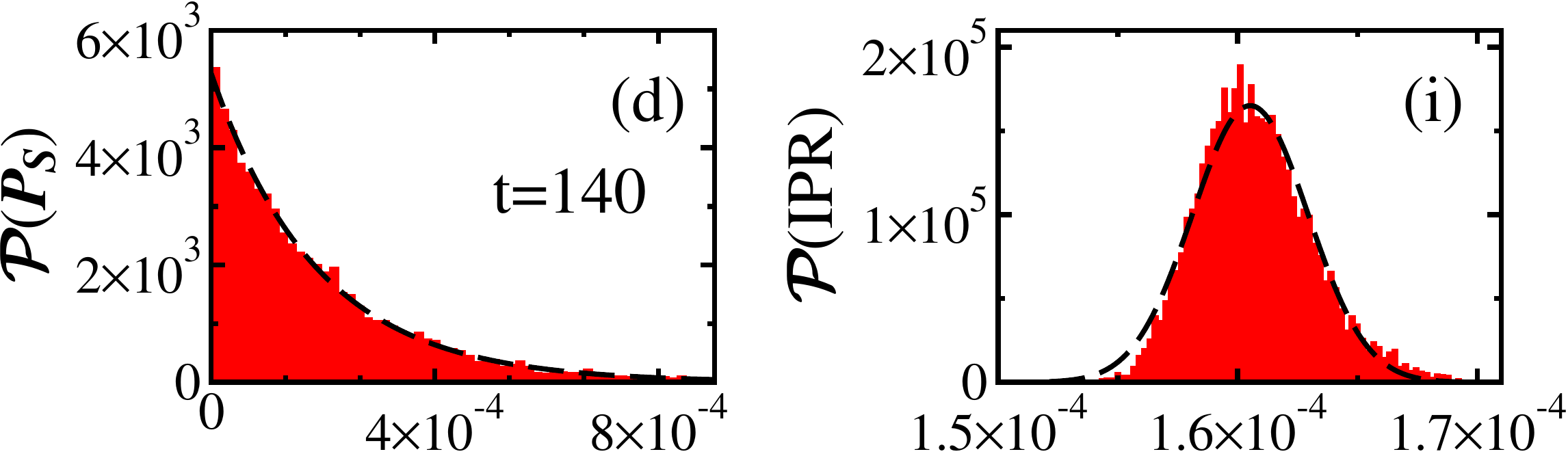}\\
\includegraphics*[width=0.41\textwidth]{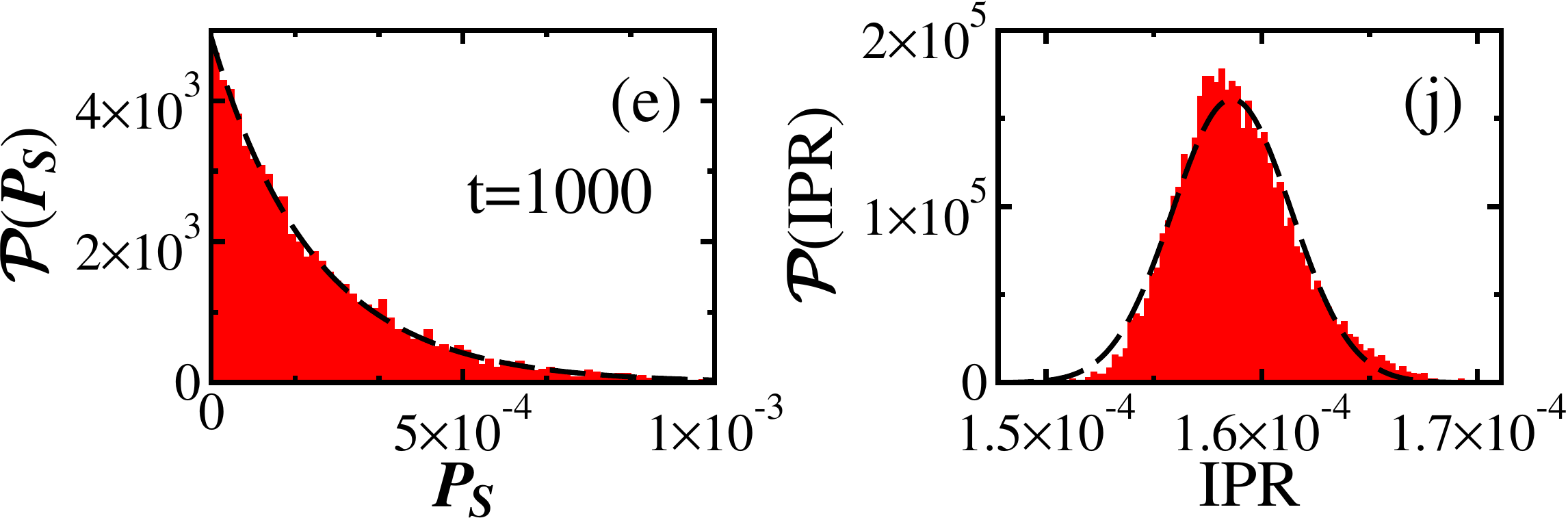}
\caption{Distributions of the survival probability (left) and inverse participation ratio (right) for the spin model at  times indicated in the panels. Dashed lines in (d) and (e) are the exponential distribution with rate parameter given by $1/\langle P_S(t) \rangle$, and in (i) and (j) they are the Gaussian distribution with the mean and variance obtained numerically.}
\label{fig:interm}       
\end{figure}

As time increases, the Gaussian distribution that $P_S(t)$ shows for the GOE model at short times becomes gradually more skewed until an exponential distribution emerges. For the spin model, the bands found in the distribution at short times [Fig.~\ref{fig:interm}~(a)] broaden  and simultaneously become more skewed [Fig.~\ref{fig:interm}~(b)] until the distribution becomes exponential as well [Figs.~\ref{fig:interm}~(d) and~\ref{fig:interm}~(e)]. 

Notice that for both models, the exponential distribution is seen even before $t_{\rm R}$. It starts taking shape already in the interval of the power-law decay [Fig.~\ref{fig:interm}~(c)] and it becomes clearly exponential at $t\sim t_{\rm{Th}}$ when the spectral correlations get manifested in the dynamics and the correlation hole develops  [Figs.~\ref{fig:interm}~(d) and~\ref{fig:interm}~(e)]. 

For $t\geq t_{\rm Th}$,  the rate parameter of the exponential distribution is given by $1/\langle P_S(t) \rangle $, as shown with a dashed line in Figs.~\ref{fig:interm}~(d) and~\ref{fig:interm}~(e). It is only for $t>t_{\rm R}$ that $1/\langle P_S(t) \rangle \sim 1/\sum_{\alpha} |C_{\alpha}^{\text{ini}}|^4$ and we recover the curve from Fig.~\ref{fig:long}~(b).
The fact that we have an exponential distribution for $P_S(t)$, with mean equal to the dispersion during the entire duration of the correlation hole $( t_{\rm Th} \leq t \leq t_{\rm R})$ implies that both $\langle P_S(t) \rangle$ and $\sigma_{P_S}(t)$ decrease below their saturation values and that ${\cal R}_{P_S}(t) \sim 1$ for any time $t \geq t_{\rm Th}$, as we indeed see numerically in Figs.~\ref{fig01}~(c) and~\ref{fig01}~(d). 

We notice that  for an integrable model, where the correlation hole does not exist, one should not expect an exponential distribution for $P_S(t)$ before saturation, that is, for $t_{\rm Th} \leq t \leq t_{\rm R}$. However, as discussed below Eq.~(\ref{eq:help}), it should emerge for $t>t_{\rm R}$. The analysis of how the distribution of the survival probability may serve as an indicator of quantum chaos is a subject worth further studies.

\subsection{Inverse participation ratio}

The distribution of ${\rm IPR}(t)$ for the GOE model is throughout Gaussian, although some level of skewness and kurtosis larger than 3 are seen for times where $\langle {\rm IPR}(t) \rangle $ oscillates, which corresponds to the power-law region of the survival probability. The width of the distribution depends on the dimension of the GOE matrix. At short times, the variance is related with the distribution of $\Gamma^2$, so it increases as the matrix grows, while at long times, the variance is related with the distributions of the components  $C_{\alpha}^{n}$, so it decreases as $D$ grows. We therefore have a Gaussian distribution that shrinks as time grows. The fact that the distribution is Gaussian at short and long times, but self-averaging holds at long times only, reiterates our claim that there is not a one to one correspondence between the shape of the distribution and the presence of self-averaging.

The distribution for ${\rm IPR}(t)$ for the spin model is hybrid. It starts similar to the distribution for the survival probability of the spin model [Figs.~\ref{fig:interm}~(f) and~\ref{fig:interm}~(g)], but it later acquires a shape equivalent to the distribution of ${\rm IPR}(t)$ for the GOE model [Figs.~\ref{fig:interm}~(i) and~\ref{fig:interm}~(j)].

\section{Inferring self-averaging behaviors from distributions}
\label{sec:local}

The main purpose of this section is to show how we can use the distribution of one quantity to assist determining the self-averaging behavior of another related quantity. But before that, we summarize the self-averaging behavior and the shapes of the distributions of the two experimental local quantities, the connected spin-spin correlation function and the spin autocorrelation function evolved under the spin model.

\subsection{Distributions of local quantities}
Both quantities $C$ and $I$ are self-averaging up to the correlation hole. The connected spin-spin correlation function does not detect the hole and remains self-averaging at all times~\cite{Schiulaz2020}. In contrast, the spin autocorrelation function exhibits a correlation hole and stops being self-averaging beyond its minimum value. 

Similarly to the survival probability and the inverse participation ratio, the distributions of the values of the two local quantities at short times also reflect the distribution of $\Gamma^2$. They exhibit fragmented structures similar to those in Figs.~\ref{fig:short}~(b) and~\ref{fig:short}~(d). However, the main difference between the global and local quantities at short times is that $P_S$ and ${\rm IPR}$ are not self-averaging, while the local quantities are, because they have an explicit dependence on the system size in the denominator~\cite{Schiulaz2020}, 
\be
I(t) \sim 1 - 4\frac{\Gamma^2 t^2}{L} ,
\ee
so
\be
{\cal R}_{I}(t) \sim 16\frac{\sigma_{\Gamma^2 }^2 t^4}{L^2} ,
\ee
which decreases with $L$.

As time grows, the distributions for the connected spin-spin correlation function and for the spin autocorrelation function progress  in a way similar to the distribution for the inverse participation ratio shown in Fig.~\ref{fig:interm}, that is, from a fragmented structure at short times to a Gaussian shape at long times.

Even though both quantities show Gaussian distributions at long times, $C$ is strongly self-averaging, with ${\cal R}_{\rm C}(t>t_{\rm R})$ decreasing exponentially as $L$ increases~\cite{Schiulaz2020}, while $I$ is non-self-averaging at long times. This is what we observe by studying system sizes with $L\leq 18$, although one cannot rule out the possibility that  this behavior might change for much larger $L$'s. Based on the results at hand, the fact that both quantities exhibit a Gaussian distribution makes us conclude that there is no direct connection between self-averaging for $t>t_{\text{R}}$ and a Gaussian distribution.

\subsection{Semianalytical results for self-averaging}

The spin autocorrelation function can reach negative values at long times, which could suggest that ${\cal R}_{I}(t)$ increases with $L$ just because $\langle I(t) \rangle$ gets very close to zero. This motivates us to study also the self-averaging behavior of $|I(t)|$ and $I^2(t)$.

\begin{figure}[ht]
\includegraphics*[width=0.45\textwidth]{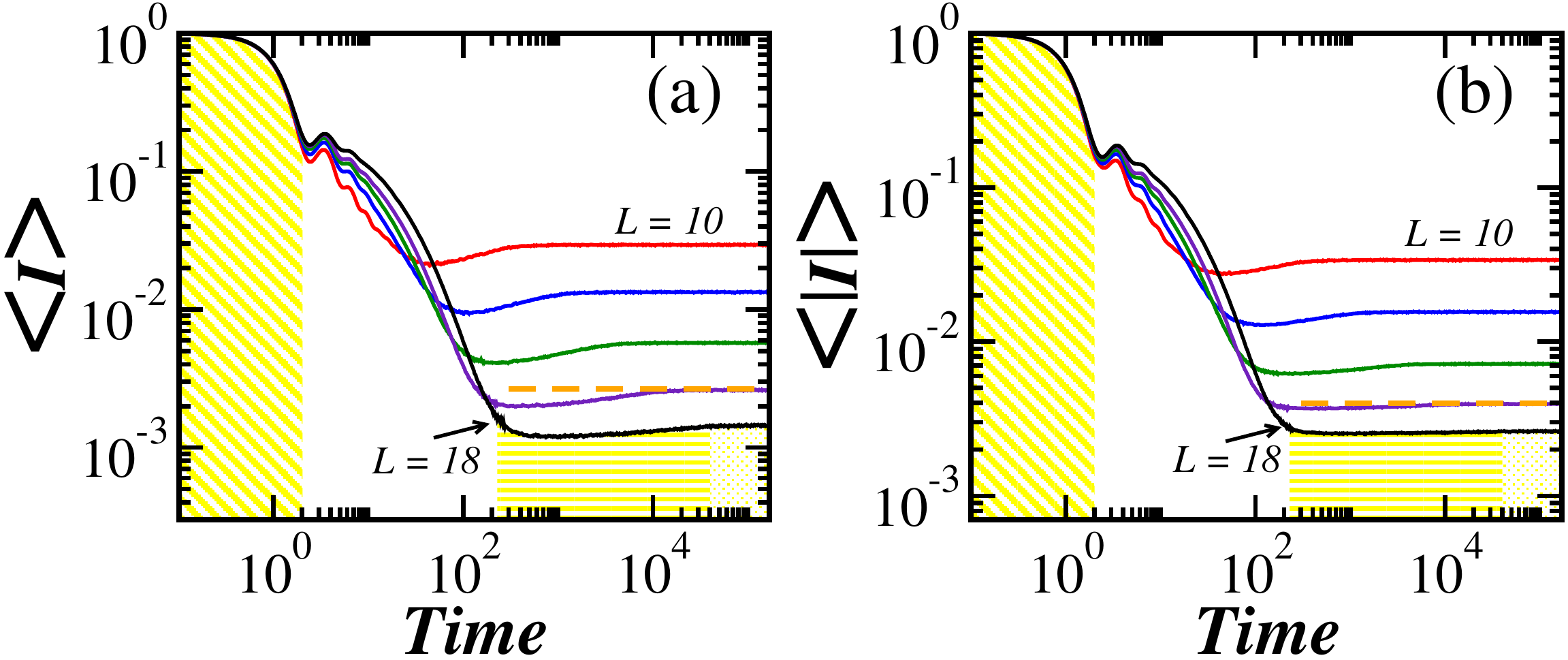}\\
\vspace{0.2cm}
\includegraphics*[width=0.3\textwidth]{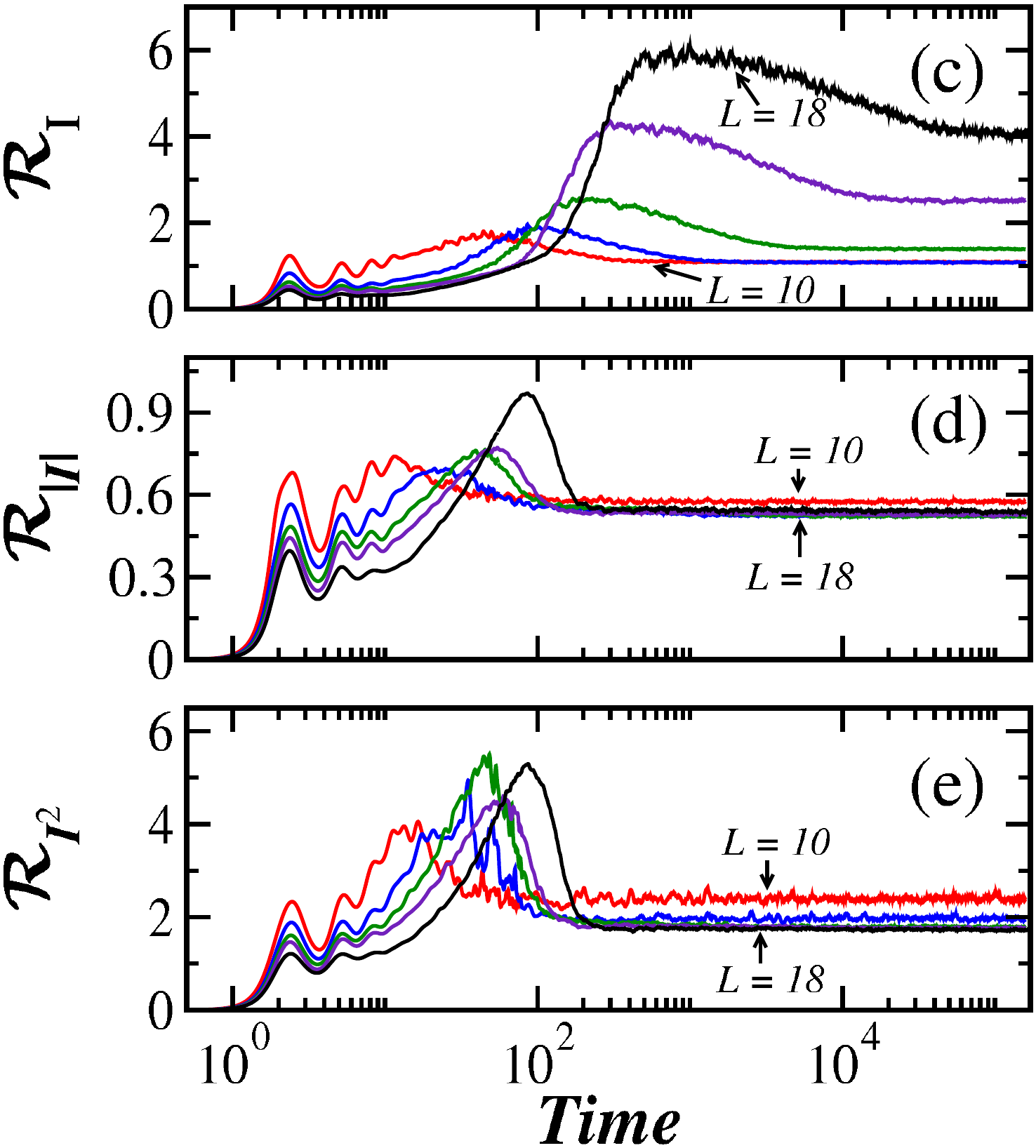}\\
\hspace{0.6 cm}
\includegraphics*[width=0.2\textwidth]{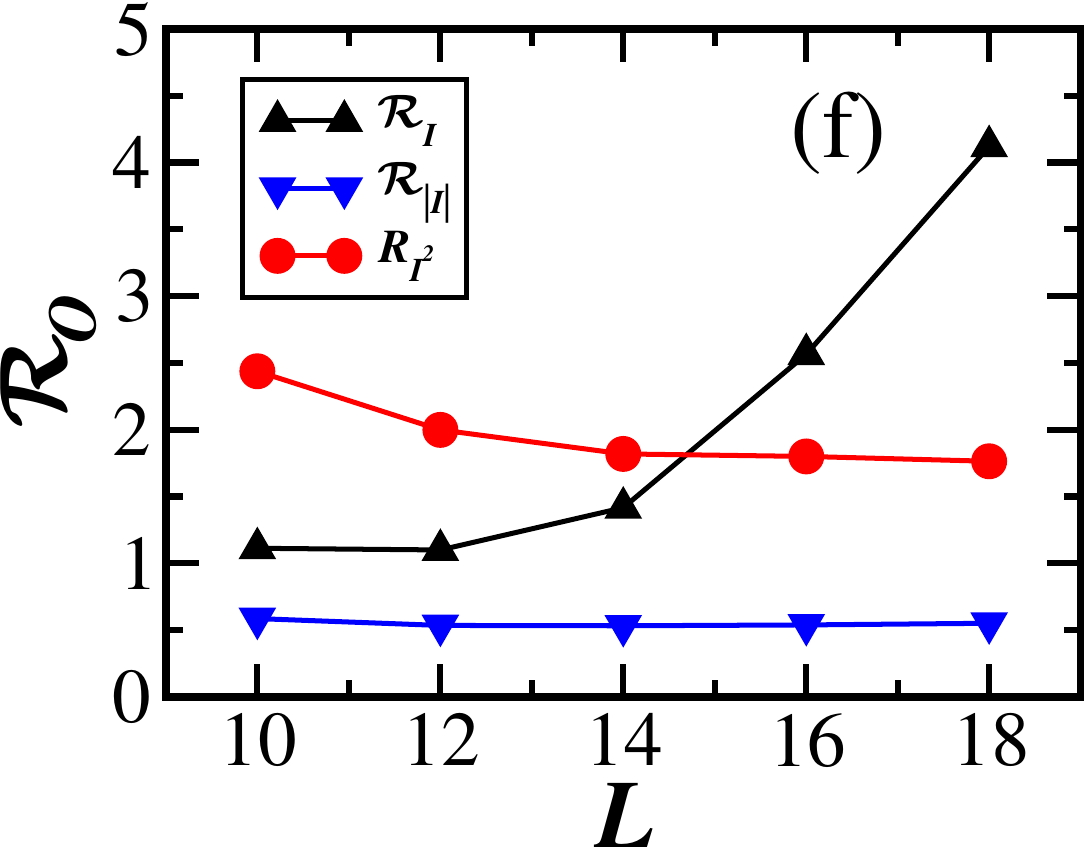}
\caption{Evolution of the mean of the spin autocorrelation function (a) and its relative variance (c), the mean of the absolute value of the spin autocorrelation function (b) and its relative variance (d), the relative variance of the square of the spin autocorrelation function (e), and the relative variance of the three quantities for $t>t_{\text{R}}$ vs. $L$ (f). All panels are for the chaotic disordered spin model. In panels (a, b) the four time intervals identified in the evolution of the survival probability are indicated, and  the horizontal dashed line marks the saturation value for $L=16$. In all panels: Average over $10^4$ data and (f) includes also an average for 100 different instants of times.}
\label{fig:imb}       
\end{figure}

In Figs.~\ref{fig:imb}~(a) and~\ref{fig:imb}~(b), we compare the results for the mean of the spin autocorrelation function and for the mean of its absolute value. The correlation hole is less evident for $\langle |I(t)| \rangle$ and for $\langle I^2(t) \rangle$ (this one is not shown) than for $\langle I(t) \rangle$, but it is still present. For the three quantities, however, the ratio between the saturation point and the minimum of the hole decreases as $L$ increases, which contrasts with the survival probability, where the ratio is constant.

As expected for local quantities, the three observables are self-averaging at short times, with ${\cal R}_{I, |I|,I^2}(t)$ decreasing as $L$ increases [see Figs.~\ref{fig:imb}~(c), \ref{fig:imb}~(d) and~\ref{fig:imb}~(e)]. For $t \sim t_{\rm Th}$,  the curves cross. Beyond this point, for $t > t_{\rm Th}$, the behavior of ${\cal R}_{I}(t)$, ${\cal R}_{|I|}(t)$, and ${\cal R}_{I^2}(t)$ differ. ${\cal R}_{I}(t)$ increases with system size, confirming the non-self-averaging behavior mentioned above, while the curves for ${\cal R}_{I^2}(t)$ cross once again, recovering self-averaging at very long times. The results for ${\cal R}_{|I|}(t)$, however, are much less conclusive. Excluding $L=10$, which is very small, the curves for $t> t_{\text{Th}}$ seem to reach a nearly constant value independent of $L$, as shown also in the scaling analysis in Fig.~\ref{fig:imb}~(f).  This suggests lack of self-averaging, but how can we better convinced of it with the system sizes that we have access to?  

Our strategy to circumvent the limited system sizes available is to use the numerical results for $I(t)$ to infer the self-averaging behavior of $|I(t)|$, as we explain next. 

We verified that distribution for $|I(t>t_{\rm R})|$ is a folded Gaussian, which further supports that the distribution for $I(t>t_{\rm R})$ is indeed Gaussian. 
Both the standard deviation and the mean of $I(t)$ for $t>t_{\rm R}$ decrease as the system size increases. The exponents $s$ and $m$ in $\sigma_I(t>t_{\rm R}) \propto L^{-s}$ and $\langle I(t>t_{\rm R}) \rangle \propto L^{-m}$ can be obtained numerically. We find that $m>s$. With this information, we can compute the mean and the variance of the folded Gaussian distribution for $|I(t)|$ using
\ba
\langle |I| \rangle &=& \sqrt{\frac{2}{\pi}} \,\sigma_I \exp \left( -\frac{\langle I \rangle^2}{2 \sigma^2_I} \right)
+ \langle I \rangle \,\text{erf} \left( \frac{\langle I \rangle}{\sigma_I} \right), 
\nonumber \\
 \sigma^2_{|I|} &=&  \langle I \rangle^2+\sigma^2_{I} - \langle |I| \rangle^2.
\nonumber 
\ea
For large $L$, we find that
\ba
\langle |I| \rangle & 
\rightarrow &  L^{-s} \sqrt{\frac{2}{\pi}} ,
 \\
 \sigma^2_{|I|} &  
 \rightarrow &
L^{-2s}  \left(  1-\frac{2}{\pi}  \right) ,
\ea
which implies that the relative variance goes asymptotically to a constant,
\be
{\cal R}_{| I |}(t>t_{\rm R})  \rightarrow \frac{\pi - 2 }{2} \sim 0.57.
\ee
This value is indeed very close to what we have in Fig.~\ref{fig:imb}~(f), but the semianalytical strategy described above provides a much stronger evidence that $| I (t)|$ is non-self-averaging at long times than what we can conclude from the numerical results in Fig.~\ref{fig:imb}~(f). 

\section{Conclusions}
\label{sec:conclusions}

We investigated the distributions over disorder realizations of different quantities and at various timescales of the evolution of a realistic chaotic spin model, from very short times up to equilibration. We compared these distributions with the quantities' self-averaging properties. The distributions for the global quantities --- the survival probability and the inverse participation ratio --- were contrasted also with those for the GOE model. The results for the two models are comparable at long times, but not at short times.

At long times, the distribution of the survival probability for the GOE and for the chaotic spin model is exponential, which accounts for the lack of self-averaging of this quantity. The exponential shape emerges as soon as the dynamics detect the spectral correlations typical of chaotic systems.

At long times, the distribution of the inverse participation ratio and also of the local quantities --- the spin-spin correlation function and the spin autocorrelation function --- are Gaussian. The fact that the first two are self-averaging, while the spin autocorrelation function is not, demonstrates that there is no direct relationship between the presence of self-averaging and the onset of a Gaussian distribution.

We also studied the absolute value and the square of the spin autocorrelation function, $|I(t)|$ and $I^2(t)$. The evolution of their mean values shows features similar to those observed for $\langle I(t) \rangle$, but their self-averaging behaviors differ. Based on the system sizes available, we conclude that at long times the spin autocorrelation function is non-self-averaging, while $I^2(t)$ is. The numerical scaling analysis of the relative variance of $|I(t)|$ is less conclusive. 

A main result of this work is to show that knowledge of the distribution of one quantity may be used to uncover the self-averaging behavior of another related quantity. This is what we achieved using $I(t)$ and $|I(t)|$ as an example. Starting with the Gaussian distribution and non-self-averaging behavior of $I(t)$ at long times, we showed semianalytically that the relative variance of $|I(t)|$ for times $t>t_{\rm{R}}$ goes asymptotically to a constant as $L$ increases, concluding in this way that $|I(t)|$ is non-self-averaging at long times. This strategy circumvents the limitations of the numerical scaling analysis, for which few system sizes can be accessed.

\begin{acknowledgments}
We are grateful to Mauro Schiulaz for various discussions during the beginning of this project. E.J.T.-H. and I.V.-F. acknowledge funding from VIEP-BUAP (Grant Nos. MEBJ-EXC19-G and No. LUAGEXC19-G), Mexico. They are also grateful to LNS-BUAP for allowing use of their supercomputing facility. L.F.S. is supported by the NSF Grant No.~DMR-1936006. 
\end{acknowledgments}

%

%

\end{document}